\documentclass[12pt,preprint]{article}

\usepackage{graphicx}

\usepackage{amsmath}

\renewcommand{\eqref}[1]{(\ref{#1})}
\newcommand{\etal}{\emph{et al.}}

\begin{document}
\begin{titlepage}
\mbox{}
\vskip3cm
\centerline{\Large\bf \sf Interaction between directional epistasis and
  average mutational effects}
\vskip 0.5cm
\vskip 0.25in
\centerline{\large \bf \sf Claus O. Wilke$^\ast$ and Christoph Adami$^{\ast\dagger}$}
\vskip 0.25in
\centerline{\it $^\ast$Digital Life Laboratory 136-93}
\centerline{\it California Institute of Technology, Pasadena, CA 91125}
\vskip 0.25in
\centerline{\it $^\dagger$Jet Propulsion Laboratory MS 126-347}
\centerline{\it California Institute of Technology, Pasadena, CA 91109}
\vskip 2cm

\end{titlepage}
\noindent
{\bf Summary: We investigate the relationship between the average fitness
  decay due to single mutations and the strength of epistatic interactions in
  genetic sequences. We observe that epistatic interactions between mutations
  are correlated to the average fitness decay, both in RNA secondary structure
  prediction as well as in digital organisms replicating in silico.  This
  correlation implies that during adaptation, epistasis and average mutational
  effect cannot be optimized independently. In experiments with RNA sequences
  evolving on a neutral network, the selective pressure to decrease the
  mutational load then leads to a reduction of the amount of sequences with
  strong antagonistic interactions between deleterious mutations in the
  population.}

\bigskip

\noindent
Keywords: epistasis, neutrality, RNA secondary structure folding, digital organisms

\section{Introduction}

A thorough understanding of epistatic interactions of mutations in genomes
is becoming more and more crucial to many areas in population genetics
and evolutionary biology. Epistasis affects linkage
disequilibria~(Charlesworth 1976, Barton 1995), robustness to
mutations~(Lenski \etal{} 1999) or canalization (Nowak \etal\ 1997, Wagner
\etal\ 1997, Rice 1998, Ancel \& Fontana 2000,
for reviews, see Scharloo 1991, Gibson \&
Wagner 2000), as well
as theories on the maintenance of
sex (Kondrashov 1982, Kondrashov 1988, West \etal 1999).
The sign of epistatic effects, that is, whether deleterious mutations are
reinforcing (synergistic epistasis) or mitigating (antagonistic epistasis)
also influences whether or not deleterious mutations can accumulate in the
genome via Muller's ratchet (Muller 1964, Felsenstein 1974).  The consensus
seems to be that synergistic epistasis can prevent the accumulation of
mutations (Crow \& Kimura 1978, Kondrashov 1994, but see Butcher 1995 for a
dissenting view). On the other hand, the observation of pervasive compensatory
mutations (Moore \etal{} 2000), which also render the ratchet powerless,
indicates epistasis, but not its sign.

While the genomes of a number of organisms have been examined for signs of
epistasis (de Visser \etal{} 1997a, 1997b, de Visser \& Hoekstra 1998, Elena
\& Lenski 1997, Elena 1999), no general trend can be discerned except to say
that interactions between mutations are frequent and of both signs, and that
weak synergistic epistasis seems to prevail in eukaryotic genomes while viral
and prokaryotic genomes show no net tendency in either direction. Experiments
to measure epistatic interactions are difficult and usually yield results of
weak statistical significance (West \etal{} 1998).  Consequently, even
epistasis of considerable strength can conceivably be missed {\it in vitro} or
{\it in vivo}. Here, we investigate deleterious mutations {\it in silico}, and
study a fact which has not received much attention in the population genetics
literature. Namely, epistasis is closely related to the geometry of the
phenotype space (Rice 1998), which leads to interesting relations between
epistasis and the effects of single mutations. Wagner \etal{} (1998) showed in
a two-locus, two-allele model that the average effect of a single mutation is
correlated to the degree of interaction between loci, and supported their
theoretical model with QTL data on body weight in mice. We demonstrate here a
similar relation for a measure of epistasis which is more appropriate for high
dimensional sequence spaces. We give both theoretical and experimental
evidence that the strength of directional epistasis is correlated with the
average deleterious effect of a single mutation. As a corollary of this
observation, we argue that in situations in which there is a selective
pressure to reduce the average deleterious effect, this correlation leads to a
reduction of the number of genomes with strong antagonistic effects in a
population.

\section{Neutrality and Epistasis}

It is a common observation that the average fitness at a mutational distance
$n$ from a given reference sequence decays roughly exponentially with $n$ (see,
e.g., de Visser \etal\ 1997b, Elena \& Lenski 1997, Lenski \etal\ 1999, and
also our results on RNA sequences below). The simplest explanation for a
\emph{perfect} exponential is a multiplicative landscape in which every
mutation diminishes the fitness independently by the same factor $(1-s)$.
Here, however, we focus on fitness landscapes with a considerable amount of
lethal mutations, in which case a branching process in the high-dimensional
sequence space is a more appropriate model: if every viable sequence has a
probability of $1-s$ to remain viable after suffering a single point mutation,
then the mean fitness will decay as (neglecting fitness differences in the
viable sequences)
\begin{equation}  \label{eq:exp-fitness-decay}
  w(n)=(1-s)^n\equiv e^{-\alpha n}\,,
\end{equation}
where we have defined $\alpha=-\ln(1-s)$, and assuming that the fitness of our
reference sequence at $n=0$ is $w(0)=1$. Both explanations for the exponential
decay have in common that the effects of subsequent mutations are independent,
i.e., there exists no epistasis. If there are some interactions between
mutations, we will still observe an exponential decay if antagonistic and
synergistic interactions occur in the same proportion. If however there exists
a bias towards either antagonistic or synergistic epistatic interactions
(directional epistasis), then this bias will naturally appear as deviations
from the exponential decay in Eq.~\eqref{eq:exp-fitness-decay}.
While such deviations have previously been
indicated by adding a term quadratic in $n$ to the exponent of
Eq.~\eqref{eq:exp-fitness-decay} (Crow 1970, Charlesworth 1990, de Visser
\etal\ 1997b, Elena \& Lenski 1997), such a parameterization
becomes troublesome at larger $n$, because $w(n)$ could increase beyond the
fitness of the reference sequence (for a positive coefficient in the quadratic
term). This is avoided by the ansatz (Lenski \etal{} 1999)
\begin{equation}\label{eq:two-param-fitness-decay}
w(n)=e^{-\alpha n^\beta}\;,
\end{equation}
where $\beta=1$ means there is no bias towards either form of epistatic
interactions. A $\beta>1$ indicates synergistic mutations prevail (mutations
that are on average ``worth'' more than one independent hit), while $\beta<1$
reflects a bias towards antagonistic mutations (mutations whose ``damage'' is
less then one independent mutation).\footnote{Note that in the earlier works
  where a quadratic term was used, the distinction between the different types
  of directional epistasis depended on whether $\beta$ was larger or smaller
  than 0, rather than 1.}  Since expression \eqref{eq:two-param-fitness-decay}
depends only on two parameters, deviations from that form may arise when $n$
grows large.

Naively, one might assume that the decay parameter $\alpha$ and the epistasis
parameter $\beta$ are independent. Instead, we shall see that environments
with strong selection force a trade-off between $\alpha$ and $\beta$, so that
one can only be optimized at the expense of the other. The reasoning is as
follows. In a strongly selective environment mutations can be classified as
either neutral or lethal, and $w(n)$ can be thought of as the fraction of
neutral sequences in genetic space at mutational distance $n$. In particular,
the \emph{neutrality} $\nu$ of a sequence (the number of sequences at Hamming
distance 1 with fitness 1)
is related to the decay parameter by $\nu = \ell(D-1) e^{-\alpha}$, where
$\ell$ is the length of the sequence, and $D$ is the number of monomers. If
all sequences in genetic space have the same $\nu$, it follows that $\beta=1$.
A deviation from $\beta=1$ implies that some sequences have more or fewer
neutral neighbors than others, giving rise to a correlation between $\alpha$
and $\beta$. For a viable sequence with lower than average neutrality (higher
than average $\alpha$), there are comparatively fewer sequences close-by than
there are far away, such that this sequence will have a small $\beta$.
Conversely, a sequence with a high neutrality (small $\alpha$) will have
comparatively more sequences close by, and $\beta$ will be larger. We can make
this argument more formal with a simple ``conservation law'', which only
reflects that the total number of neutral sequences in genetic space is
constant.  Since for polymers of fixed length $\ell$ made from $D$ monomers
there are $\binom \ell n (D-1)^n$ possible $n$-mutants, we must have
\begin{equation}
\sum_{n=1}^{\ell} w(n)\,\binom \ell n (D-1)^n = N_\nu\;, \label{sumrule}
\end{equation}
where $N_\nu$ is the total number of neutral mutants of this wild type.
Inserting $w(n)$ from Eq.~\eqref{eq:two-param-fitness-decay} yields an
implicit relation between $\alpha$ and $\beta$. Indeed, for two decay
functions of two different reference sequences, with different parameters
$\alpha$, the only way in which the sum can yield the quantity $N_\nu$ (which
is independent of the respective reference sequence) is that the two decay
functions must have different parameters $\beta$ as well.  However, the
implicit relation depends on the ansatz
Eq.~\eqref{eq:two-param-fitness-decay} being correct for all $n$, which is not
necessarily the case. Alternatively, we may consider only sequences with up to
$d$ mutations, in which case we can write
\begin{equation}
\sum_{n=1}^{d} w(n)\,\binom \ell n (D-1)^n = N_{\nu,d}\,, \label{sumrule2}
\end{equation}
where $N_{\nu,d}$ is the number of neutral sequences in a sphere of radius $d$
around the reference sequence with $w(0)=1$.  $N_{\nu,d}$ depends on the
particular reference sequence chosen, and therefore cannot be regarded a
constant. However, for $d$ not too small, we may replace it by its average
$\langle N_{\nu,d}\rangle$ over all viable reference sequences. If we
take into account a sufficiently large region of genotype space, we should
find roughly the same number of neutral mutants for each viable sequence,
since, as $d$ approaches $\ell$, the quantity $N_{\nu,d}$ approaches $N_{\nu}$,
which in turn is independent of the reference sequence.  Hence,
Eq.~\eqref{sumrule2} predicts a similar relation between $\alpha$ and $\beta$,
and the two different predictions approach each other as $d\rightarrow l$.
Predictions based on Eqs.~\eqref{sumrule} and~\eqref{sumrule2} are used below
to compare to our empirical results.

Although the above argument strictly holds only under the assumption that
mutations are either neutral or lethal, it is not unreasonable to assume that
a similar (possibly weaker) correlation between $\alpha$ and $\beta$ exists
also in more general cases, where slightly deleterious or even advantageous
mutations are possible. In that case, under the presence of epistasis, there
will still be regions in genotype space in which the number of
less-deleterious mutations is higher, and other regions in which it is lower
than average. The decay function $w(n)$ of a sequence from a region that is
rich in non-lethal mutations would have a lower $\alpha$, but would be
inevitably more synergistic than the decay function of a sequence from a
region poor in non-lethal mutations. Our results with digital organisms (see
below) support this reasoning.

\section{Experimental Evidence}

Accurate data for the decay parameter $\alpha$ and the epistasis parameter
$\beta$ for biological organisms are rare, which makes our hypothesis
difficult to test. A few well-studied systems have emerged
which are accessible {\it in silico}.

\begin{figure}[tb]
\centerline{\includegraphics[width=9cm]{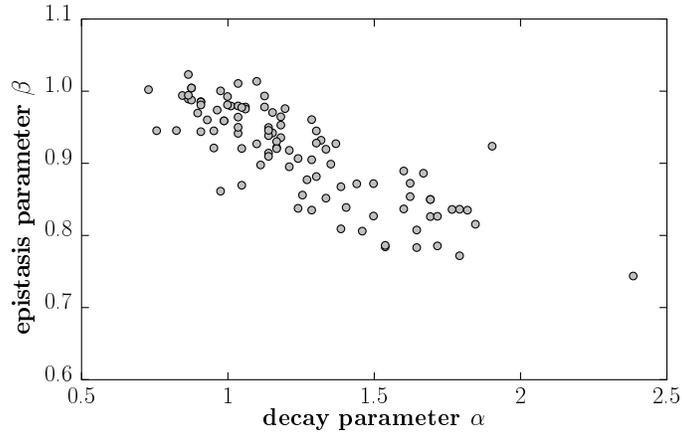}}
\caption{\label{fig:RNA-l76}Epistasis parameter $\beta$ versus the decay
 parameter $\alpha$ in random RNA sequences of length $\ell=76$.}
\end{figure}

We studied RNA secondary structure prediction using the Vienna RNA package,
version 1.3.1., with the default setup (Hofacker \etal{} 1994). We calculated
the decay of the average number of neutral folds as a function of the Hamming
distance for 100 random RNA sequences of length $\ell=76$. The parameters
$\alpha$ and $\beta$ were determined as follows. We obtained
$\alpha$ exactly from the fraction of neutral one-mutants. In addition, we
sampled the function $w(n)$ for Hamming distances up to $n=8$, by calculating
the structure of up to $10^6$ random neighbors of the required Hamming
distance. The quantity $\beta$ was then determined from a nonlinear fit of
$-\alpha n^\beta$ to the logarithm of $w(n)$.  A plot of $\beta$ versus
$\alpha$ (Fig.~\ref{fig:RNA-l76}) shows a significant correlation, with a
correlation coefficient of $r=-0.817$ ($p<0.01$).

\begin{figure}[tb]
  \centerline{\includegraphics[width=9cm]{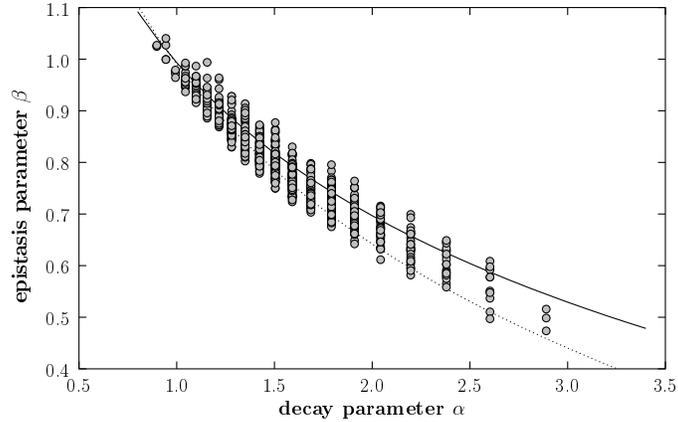}}
\caption{\label{fig:beta-vs-alpha}Epistasis parameter $\beta$ versus the
  decay parameter $\alpha$ for sequences on the larger of the two neutral
  networks from (van Nimwegen \etal{} 1999). The solid line represents the
  prediction that follows from Eq.~\eqref{sumrule}, and the dotted line
  represents the prediction from Eq.~\eqref{sumrule2}, with $d=8$.}
\end{figure}

According to Eqs.~\eqref{sumrule} and~\eqref{sumrule2}, we can predict the
relationship between $\alpha$ and $\beta$ if we compare the decay functions
of sequences that are mutually neutral. For RNA folding, this means we have to
determine $\alpha$ and $\beta$ for a set of sequences that fold into the same
structure.  We performed experiments with the RNA sequences of length
$\ell=18$ used by van Nimwegen \etal{} (1999)
(for these sequences we altered the
default setup by setting the free energies of dangling ends to zero.)  For the
particular case that all bonds are of the purine-pyrimidine type (G-C, G-U,
A-U), two separate neutral networks (a neutral network is a network of neutral
genotypes connected to each other by one-point mutations) were found by van
Nimwegen \etal{}, consisting of 51,028 and 5,169 sequences, respectively.  For
each such set of neutral sequences, Eq.~\eqref{sumrule} predicts the
correlation without free parameter, as long as the number of {\it all} neutral
sequences $N_\nu$ is known.  In order to estimate $N_\nu$, we generated $10^8$
random sequences of length $\ell=18$, of which 10,961 sequences folded
correctly.  From this, we estimated $N_\nu=7.5\times 10^6$ neutral sequences
out of the total $6.9\times 10^{10}$ sequences of length $\ell=18$. Using this
number, Eq.~\eqref{sumrule} predicts the solid line in
Fig.~\ref{fig:beta-vs-alpha}, which describes the correlation well. The
approach based on averaging over \emph{local} neutral sequences around the
reference sequence [Eq.~\eqref{sumrule2}] gives rise to the dotted line in
Fig.~\ref{fig:beta-vs-alpha} and shows that this too predicts the correlation
fairly well (for this second approach, we used $d=8$, and from Monte Carlo
simulations for 1000 reference sequences, we obtained $\langle N_{\nu,8}\rangle=2.4\times 10^5$).

As our second test case, we analyzed the correlation between $\alpha$ and
$\beta$ in digital organisms (Adami 1998, Adami \etal{} 2000). Digital
organisms are self-replicating computer programs that mutate and evolve. Their
fitness is determined by the ratio of their CPU speed and their gestation
time. The latter is given by the number of instructions that have to be
executed in order to generate a fully functional offspring. The CPU speed
increases when the digital organisms perform logical operations on numbers
that they can acquire from their environment. The gestation time, on the other
hand, increases either if the organisms employ less efficient mechansims of
self-replication, or if they accumulate instructions that are involved in the
completion of the above mentioned logical operations. The digital organisms
with the highest fitness thus have a very efficient copy mechanism, and
perform a large number of logical operations with a comparatively small number
of additional instructions.  Lenski \etal{} (1999) measured the decay of the
mean fitness as a function of the number of mutations accumulated in such
digital organisms, and obtained $\alpha$ and $\beta$ from a fit of
Eq.~\eqref{eq:two-param-fitness-decay} to the measured decay functions. They
studied 174 different genomes consisting of two groups of 87 genomes each. The
first group of organisms evolved in 87 independent experiments in a complex
environment, while the second group was obtained by transfering these
organisms to an environment which favoured simple genomes, and allowing the
organisms to adapt to this more simple environment.

A statistical analysis revealed a significant correlation between the decay
parameter $\alpha$ and the parameter of directional epistasis $\beta$ for both
the complex and the simple digital organisms (Table~\ref{tab:avida-data}).
However, in addition to said correlation, we found a correlation between the
$\alpha$ and both the genome length $\ell$ and the log fitness for complex and
simple organisms, as well as a correlation between $\beta$ and $\ell$ in
the case of the complex organisms. Hence, for the complex organisms, we cannot
rule out the possibility that the correlation between $\alpha$ and $\beta$
merely reflects an underlying correlation of both quantities with length. In
the case of the simple organisms, where we do not see a correlation between
$\beta$ and $\ell$, we can assume that the correlation between $\alpha$ and
$\beta$ is genuine. To provide further evidence, we examined a reduced data
set of all 48 simple organisms with a length between $\ell=14$ and $\ell=16$.
These 48 organisms were also of comparable fitness.  In that data set, we
found an even stronger correlation between $\alpha$ and $\beta$, while the
correlation between either of the two quantities and length or fitness was
insignificant (Table~\ref{tab:avida-data}).  It was not possible to study a
similar reduced data set for the complex organisms, because the variations in
length were too large (the length varied between $\ell=20$ and $\ell=314$
among the 87 genomes). Although the data from the digital organisms is
potentially biased because our reference sequences were evolved rather than
random as in the RNA case, we believe this bias to be insubstantial in at
least the reduced data set, which contains only organisms with very similar
phenotypes. Hence, just as for our RNA data, the data from the digital
organisms supports our hypothesis of a genuine correlation between $\alpha$
and $\beta$.

\section{Adaptation of Epistasis through Correlated Response}

The correlation between neutrality and epistasis implies that if one of them
is subject to selective pressures, the other will be affected as well due to
correlated response.  Van Nimwegen \etal{} (1999) have shown that a population
evolving on a neutral network reduces its genetic load by moving into the
regions of high neutrality in sequence space.  In particular, given a random
population of molecules on such a network, evolution tends to increase the
neutrality in the population, effectively pushing the population into the {\it
  center} of the neutral network. Because of the correlation between
neutrality and epistasis, we expect this dynamic to lead to a reduction of
antagonistic epistatic effects. To verify this hypothesis, we carried out
evolutionary experiments with the RNA sequences of length 18 from the previous
section.

\begin{figure}[tb]
\centerline{\includegraphics[width=8cm]{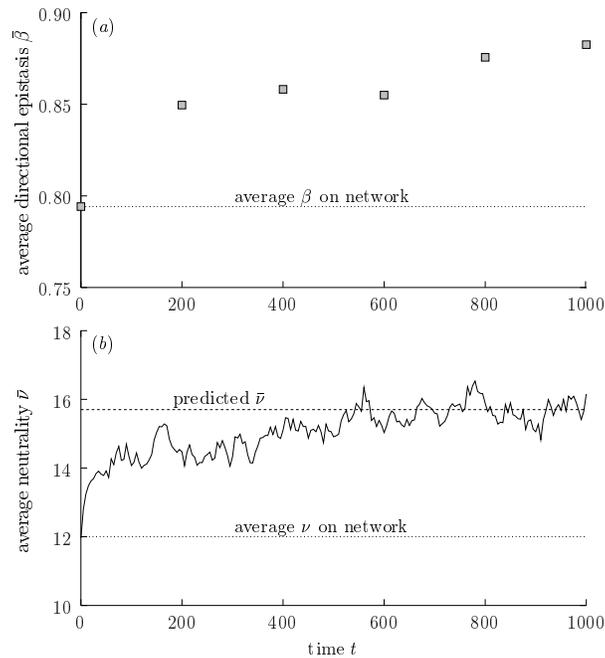}}
\caption{\label{fig:set1.run1} Evolution of neutrality and
  epistasis as a function of time (in generations). The lower graph
  shows the convergence of the population neutrality to the value
  predicted by the spectral radius of the connectivity matrix. The
  upper graph shows the change of $\beta$, averaged over the
  population, in the same run. For all data points in the upper graph,
  the standard error of the mean does not exceed the size of the
  symbols.}
\end{figure}

We performed one flow-reactor run for each of the two networks found by van
Nimwegen \etal{}, starting with an initial population of 1,000 sequences
chosen at random from the respective network. We set the replication rates
such that sequences folding into the target structure replicated on average
once per unit time, while the replication rate of all other sequences was set
to $10^{-6}$ per unit time. All sequences had a probability of $\mu=0.5$ to
suffer one random point mutation per replication event. The possibility of
several point mutations per replication event was eliminated, to guarantee
that the population could not leave the specified neutral network. The
epistasis parameter $\beta$ was determined for every sequence in the
population every two hundred generations, while the population neutrality was
monitored constantly. The population neutrality $\bar\nu$ is the average
neutrality of all sequences currently in the population. In
Fig.~\ref{fig:set1.run1}, we present the results from the run on the larger of
the two networks. The neutrality of the initial population coincides with the
network neutrality (the average neutrality of all sequences on the network),
which is to be expected for a random initial population. Over the course of
evolution, the average population neutrality rose to the predicted equilibrium
value (given by the spectral radius of the connectivity matrix, see van
Nimwegen \etal{} 1999.) As expected, the average epistasis parameter
$\bar\beta$ increased significantly as well.  Results on the second network
were qualitatively identical, with $\beta$ increasing from 0.78 to around
0.86.  Thus, antagonistic epistasis is reduced during
adaptation for reduced mutational load on a neutral network.

The above reasoning depends of course on the assumption that a population
remains on a single neutral network. In a more realistic scenario, where peaks
of different heights are present, the main effect of selection will be to
increase the fitness, rather than to increase $\alpha$. However, once a local
optimum has been reached, we can expect the dynamic described above to take
place, as long as there exists some neutrality at the local peak. If there was
a correlation between the fitness and $\alpha$ or $\beta$, then we would see
additionally a correlated response in $\alpha$ or $\beta$ as the fitness is
being maximized. Nevertheless, while the correlation between $\alpha$ and
$\beta$ reported here is a general result that follows from geometric
constraints on the landscape, no such constraints exist between the fitness
and $\alpha$ or $\beta$. The parameters $\alpha$ and $\beta$ measure the
amount and the distribution of neutral or nearly neutral sequences in the
neighborhood of a reference sequence. There is no reason why sequences with
high fitness should generally be found in regions with particularly high or
low neutrality, or with a particular distribution of the neutral sequences.
This is not to say, however, that such a correlation cannot exist in special
cases (see, e.g., our data from the complex digital organisms,
Table~\ref{tab:avida-data}, where $\alpha$ is correlated with fitness, but
$\beta$ is not).

\section{Conclusions}

Epistasis plays an important role in evolutionary theory, but remains
empirically largely unexplored. Using secondary structure prediction of RNA
sequences as well as digital organisms evolving {\it in silico}, we have
demonstrated a correlation between two important parameters of realistic
genetic fitness landscapes: the average deleterious effect of single mutations
and the strength of directional epistasis. In conjunction with the results
from Wagner \etal\ (1998) for mice, and the two different theoretical
explanations (two-locus model in the case of Wagner \etal, sequence space
based model here), we can expect this correlation to be an ubiquitious
phenomenon, present in many natural and artificial fitness landscapes.

The correlation, coupled with the
selective pressure which forces random sequences in a neutral network to
cluster in the dense areas of the network, leads to a reduction of strong
antagonistic epistasis in a population. The nature of this result is purely
geometric: as a population tries to reduce the average effect of single
mutations, the effect of multiple mutations is inevitably worsened as long as
there exist some inhomogeneities in the effect of single mutations across the
genotype space. As the result of this geometric constraint, a member of an
evolved population will have, on average, a higher $\beta$ than a random
sample of the fitness landscape would indicate.

It is well known that antagonistic epistasis favors the accumulation of
deleterious mutations as well as the operation of Muller's ratchet. Since in
such a situation sexual recombination (within a fixed environment) tends to
worsen the loss of information, recombination is unlikely to evolve or be
maintained. The mechanism described here may thus provide a path towards an
environment more conducive to the evolution of recombination.

\section{Acknowledgements}

We thank Martijn Huynen for providing access to the sequences used in (van
Nimwegen \etal{} 1999), and Walter Fontana for providing us with his flow
reactor code. We are grateful to Richard Lenski and two anonymous referees for
many useful comments and suggestions regarding the manuscript. The genomes of
digital organisms used in this study are available at\\ {\tt
  http:dllab.caltech.edu/pubs/nature99/nature.shtml}. This work was supported
by the National Science Foundation under contract No.\ DEB-9981397. Part of
this work was carried out at the Jet Propulsion Laboratory, under a contract
with the National Aeronautics and Space Administration.

\newpage

\begin{table}
\center
\begin{tabular}{|r|c|c|r|c|c|}\hline
Complex & $r$ & $p$& Simple & $r$ & $p$ \\
\hline
$\alpha$ and $\beta$ & $-0.599$ & $<0.01$
& $\alpha$ and $\beta$ & $-0.767$ & $<0.01$\\
$\alpha$ and $\ell$ & $-0.730$ & $<0.01$
& $\alpha$ and $\ell$ & $-0.502$ & $<0.01$\\
$\beta$ and $\ell$ & $0.338$ & $<0.01$
&$\beta$ and $\ell$ & $0.058$ & $>0.05$\\
$\alpha$ and $\ln w$ & $-0.333$ & $<0.01$
&$\alpha$ and $\ln w$ & $0.561$ & $<0.01$ \\
$\beta$ and $\ln w$ & $0.112$ & $>0.05$
&$\beta$ and $\ln w$ & $-0.071$ & $>0.05$ \\
\hline
Reduced  & $r$ & $p$ \\
\cline{1-3}
$\alpha$ and $\beta$ & $-0.945$ & $\ll0.01$\\
$\alpha$ and $\ell$ & $-0.101$ & $>0.05$\\
$\beta$ and $\ell$ & $-0.073$ & $>0.05$\\
$\alpha$ and $\ln w$ & $-0.0289$ & $>0.05$ \\
$\beta$ and $\ln w$ & $0.0466$ & $>0.05$\\
\cline{1-3}
\end{tabular}
\caption{\label{tab:avida-data} The correlation $r$ and $p$-value between
decay parameter $\alpha$, epistasis parameter $\beta$, length $\ell$ and the
logarithm of the fitness $\ln w$
in the data from (Lenski \etal{} 1999). The ``Complex'' and the
``Simple'' data set consist each of 87 digital organisms, the ``Reduced'' data
set consists of all 48 organisms of length
between 14 and 16 taken from the ``Simple'' data set.}
\end{table}

\end{document}